\documentclass[default]{sn-jnl}% Default
\usepackage{graphicx}%
\usepackage{multirow}%
\usepackage{amsmath,amssymb,amsfonts}%
\usepackage{relsize}
\usepackage{amsthm}%
\usepackage{mathrsfs}%
\usepackage[title]{appendix}%
\usepackage{xcolor}%
\usepackage{textcomp}%
\usepackage{manyfoot}%
\usepackage{booktabs}%
\usepackage{algorithm}%
\usepackage{algorithmicx}%
\usepackage{algpseudocode}%
\usepackage{listings}%
\usepackage{subfigure}%
\usepackage{natbib}%
\usepackage{amssymb,bbding,oplotsymbl,MnSymbol}
\definecolor{maroon}{rgb}{0.64, 0.08, 0.18}	
\theoremstyle{thmstyleone}%
\theoremstyle{thmstyletwo}%

\theoremstyle{thmstylethree}%

\begin{document}

% \title[Article Title]{A moving contact line dynamics near liquid-liquid systems}

\title[Article Title]{An experimental investigation of flow fields near a liquid-liquid moving contact line}


\author[1]{\fnm{Charul} \sur{Gupta}}\email{me18resch11001@iith.ac.in}

\author[1]{\fnm{Lakshmana D} \sur{Chandrala}}\email{lchandrala@mae.iith.ac.in}

\author*[1,2]{\fnm{Harish N} \sur{Dixit}}\email{hdixit@mae.iith.ac.in}

\affil*[1]{\orgdiv{Department of Mechanical \& Aerospace Engineering}, \orgname{Indian Institute of Technology Hyderabad}, \orgaddress{\street{Kandi}, \city{Sangareddy}, \state{Telangana}, \country{India}, \postcode{502284}}}
\affil*[2]{\orgdiv{Centre for Interdisciplinary Programs}, \orgname{Indian Institute of Technology Hyderabad}, \orgaddress{\street{Kandi}, \city{Sangareddy}, \state{Telangana}, \country{India}, \postcode{502284}}}


\abstract{
A moving contact line occurs at the intersection of an interface formed between two immiscible liquids and a solid. According to viscous theory, the flow is entirely governed by just two parameters, the viscosity ratio, $\lambda$, and the dynamic contact angle, $\theta_d$. While a majority of experimental studies on moving contact lines involve determining the relationship between the dynamic contact angle and capillary number, a few studies have focused on measuring the flow field in the vicinity of the contact line involving liquid-gas interfaces. However, none of the studies have considered liquid-liquid moving contact lines and the present study fills this vital gap. Using particle image velocimetry, we simultaneously measure the velocity field in both the liquid phases using refractive index matching techniques. The flow field obtained from experiments in both phases is directly compared against theoretical models. Measurement of interface speed reveals that material points rapidly slow down as the contact line is approached. Further, the experiments also reveal the presence of slip along the moving wall in the vicinity of the contact line suggesting a clear mechanism for how the singularity is arrested at the contact line.} 

\keywords{Dynamic contact angle, two-phase flows, interfacial flows}



\maketitle

\section{Introduction}\label{sec:intro}
The dynamics of liquid-gas moving contact line is a classical problem in fluid mechanics and has been discussed extensively in several reviews \cite{dussan1979spreading,pomeau2002recent,sui2014numerical,snoeijer2013moving,shikhmurzaev2020moving}. The problem attracted a lot of attention owing to the singularity at the moving contact line when the standard boundary condition of `no-slip' is applied on the moving wall. The presence of the singularity was first recognised by Huh \& Scriven \cite{huh1971hydrodynamic} (HS71 hereafter) wherein the authors obtained a `local' solution in the vicinity of the contact line by assuming the interface to be flat. In the viscous limit, HS71 theory predicts three different flow regimes based on the value of the viscosity ratio, $\lambda=\mu_1/\mu_2$, and the dynamic contact angle, $\theta_d$. In a recent study \cite{gupta2023experimental}, we showed that the HS71 theory quantitatively predicts the flow field away from the moving contact line, but the experiments were restricted to liquid-gas systems only. There are very few experimental studies involving a moving contact line formed in a liquid-liquid system \cite{fermigier1991experimental,zheng2021effects}, but these too are restricted to measuring the dynamic contact angle without any information about the flow field in the two liquids. There are two challenges involved in determining the flow field in liquid-liquid moving contact lines. The first is the high-resolution needed to test viscous theories which are essentially local in nature, and the second is the ability to measure flows in both fluids simultaneously.

In the present study, we conduct particle image velocimetry (PIV) experiments in a liquid-liquid moving contact line for viscosity ratio, $\lambda \geq 1$. The primary objective of the study is to measure the flow field in both liquids simultaneously using a refractive index (RI) matched PIV technique and compare the results against theoretical predictions. The secondary objective is to measure flow along the interface as well as the moving plate to determine how the singularity at the moving contact line, present in HS71 theory, is resolved. 

The experimental parameters in the present study could also be of interest to the modelling community. Numerical simulations involving moving contact lines are particularly challenging when there is a large contrast between the density and viscosity ratio. As a result, several studies especially those relying on phase-field models, have focused on systems where the viscosity ratio is close to unity \cite{yue2010sharp,sui2014numerical}. The parameters used in the present study make it now possible to validate the numerical results against the experiments quantitatively.

The paper is organized as follows: In \S \ref{sec:experimental_setup}, we discuss the experimental set-up and the parameter regimes investigated. After a brief review of relevant theoretical models in \S \ref{sec:theory_background}, the results are presented in \S \ref{sec:results}. We summarize key observations and the outcomes of the study in \S \ref{sec:conclusion}.

%%%%%%%%%%%%%%%%%%%%%%%%%%%%%%%%%%%%%%%%%%%%%%%%%%%%%%%%%%%%
%%%%%%%%%%%%%%%%%%%%%%%%%%%%%%%%%%%%%%%%%%%%%%%%%%%%%%%%%%%%
%%%%%%%%%%%%%%%%%%%%%%%%%%%%%%%%%%%%%%%%%%%%%%%%%%%%%%%%%%%%
%%%%%%%%%%%%%%%%%%%%%%%%%%%%%%%%%%%%%%%%%%%%%%%%%%%%%%%%%%%%
%%%%%%%%%%%%%%%%%%%%%%%%%%%%%%%%%%%%%%%%%%%%%%%%%%%%%%%%%%%%
%%%%%%%%%%%%%%%%%%%%%%%%%%%%%%%%%%%%%%%%%%%%%%%%%%%%%%%%%%%%
\section{Experimental setup and parameter regimes}
\label{sec:experimental_setup}

%%%%%%%%%%%%%%%%%%%%%%%
%%%%%%%%%%%%%%%%%%%%%%%
\begin{figure}
\centering
\subfigure[]{
\label{fig:experimental_setup}
\includegraphics[trim = 10mm 1mm 2mm 2mm, clip, angle=0,width=0.58\textwidth]{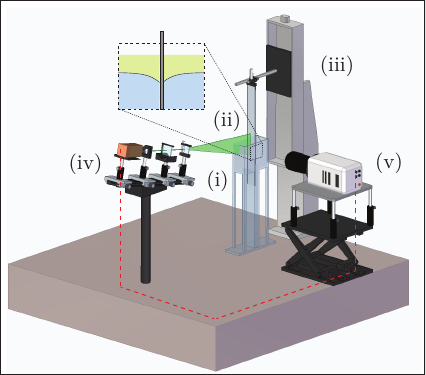}}
\hspace{-2mm}
\subfigure[]{
\label{fig:two_phase_plot}
\includegraphics[trim = 2mm -30mm 10mm 0mm, clip, angle=0,width=0.4\textwidth]{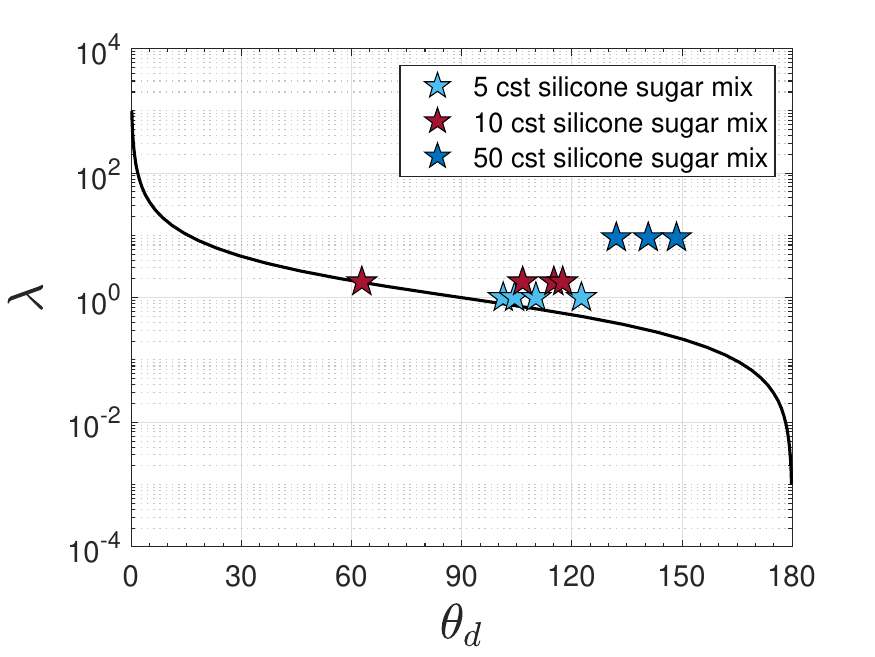}}
\caption{(a) A schematic illustration of the experimental setup and 2D PIV measurement system. (i) Rectangular tank containing two immiscible liquid phases (indicated in the inset) (ii) Glass plate (iii) Linear motorized traverse (iv) Laser along with sheet optics (v) High-speed camera with a macro lens. (b) Phase plot of viscosity ratio and dynamic contact angle plane. The black solid curve represents the theoretical critical viscosity ratio curve predicted by Huh \& Scriven \cite{huh1971hydrodynamic}. The symbols indicate the experimental data points.}
\end{figure}
%%%%%%%%%%%%%%%%%%%%%%
%%%%%%%%%%%%%%%%%%%%%%
The experimental setup shown in figure \ref{fig:experimental_setup} consists of a transparent rectangular tank of dimensions $200 \text{mm} \times100\text{mm}\times 27\text{mm}$ filled with two immiscible liquids. Standard PIV optics involving a 2W continuous diode laser and a set of a bi-concave spherical lens and plano-convex cylindrical lens generate a thin laser sheet of approximately 0.5 mm thickness. A 1 MP high-speed Photron Nova S9 camera with a macro lens captures images at frame rates of 50 to 500 fps depending upon the speed of the slide. The field of view is varied between 5.1mm $\times$ 5.1mm to 7.9mm $\times$ 7.9mm to obtain the flow fields in both phases with spatial resolution of $\approx 5 \mu$m/pixel to 8$\mu$m/pixel respectively. A traverse loaded with a vertical glass slide is immersed in the tank at a controlled speed. To keep the contact line and interface position fixed during imaging, liquid from the bottom of the tank was removed at a prescribed rate using a programmable syringe pump. Two different kinds of seeding particles $2 \mu$m for sugar-water mixture and $5\mu$m for silicone oil are utilized to capture the flow near the contact line. 

%%%%%%%%%%%%%%%%%%%%%%%%%%%%%%
%%%%%%%%%%%%%%%%%%%%%%%%%%%%%%
\begin{table}[h]
\centering % instead of \begin{center}
\caption{Properties of fluids used in experiments}
\begin{tabular}{ c @{\hskip 0.3in} c  c  c  c } 
\hline
      &  \begin{tabular}{@{}c@{}}Density \\ kg/m$^3$\end{tabular} & 
      \begin{tabular}{@{}c@{}}Viscosity \\ $10^{-3} Pa.s$\end{tabular}  & 
      \begin{tabular}{@{}c@{}}Surface tension \\ $m$N/m\end{tabular}  \\
\hline
Air & 1.207 & 0.0189 & - \\ 
Water & 1000 & 0.89 & 72 \\ 
SW-1$^*$: Sugar-water mixture & 1176 & 4.54 & 75.1 \\
SW-2$^*$: Sugar-water mixture & 1180 & 4.9 & 75.1 \\
5 cSt Silicone oil & 913 & 4.56 & 19.5\\
10 cSt Silicone oil & 940 & 8.8 & 19.5\\
50 cSt Silicone oil & 970 & 46.1 & 19.5\\
\hline
\label{tab:properties_fluids}
\end{tabular}
\end{table}
\vspace{-9mm}
\begin{flushleft}
\begin{tablenotes}
      \footnotesize
      \item $^*$SW represents refractive index (RI) matched sugar-water mixtures
    \end{tablenotes}
\end{flushleft}
%%%%%%%%%%%%%%%%%%%%%%%%%%%%%%
%%%%%%%%%%%%%%%%%%%%%%%%%%%%%%

%%%%%%%%%%%%%%%%%%%%%%%%%%%%%%
%%%%%%%%%%%%%%%%%%%%%%%%%%%%%%
\begin{table}[h]
\centering % instead of \begin{center}
\caption{Properties of liquid-liquid combinations}
\begin{tabular}{ c @{\hskip 0.3in} c  c  c  c  c c} 
\hline
      \begin{tabular}{@{}c@{}}Fluid-1\end{tabular} & 
      \begin{tabular}{@{}c@{}}Fluid-2\end{tabular}  & 
      \begin{tabular}{@{}c@{}}Density difference\\ $\Delta\rho$,  kg/m$^3$\end{tabular} &
      \begin{tabular}{@{}c@{}}Interfacial Tension\\ $\gamma$,$m$N/m\end{tabular} &
      \begin{tabular}{@{}c@{}}$\lambda = \mu_1/\mu_2$\end{tabular} &
      \begin{tabular}{@{}c@{}}$l_c = \sqrt{\frac{\gamma}{\Delta \rho g}}$\\(mm)\end{tabular} &
      \begin{tabular}{@{}c@{}}$n$\\ (RI) \end{tabular}\\
\hline
 5cSt Si oil   & SW-1 & 263  & 35.03  & 1 & 3.68 & 1.3990 \\
 10cSt Si oil  & SW-2 & 240  & 37.04  & 1.8 & 3.96 & 1.4010\\
 50cSt Si oil   & SW-2 & 210  & 36.50  & 9.4 & 4.20 & 1.4024 \\
\hline
\label{tab:properties_combinations}
\end{tabular}
\end{table}
%%%%%%%%%%%%%%%%%%%%%%%%%%%%%%
%%%%%%%%%%%%%%%%%%%%%%%%%%%%%%

In order to allow the laser sheet to pass through both liquids without any distortion or reflections from the interface, both fluids must have the same refractive index. First, polydimethylsiloxane (silicone oil) of a known viscosity is chosen and then the sugar-water mixture is prepared such that its refractive index matches that of silicone oil. Since the refractive index of the sugar-water mixture varies as the concentration of sugar changes, an appropriate concentration of sugar was chosen till the refractive index of the sugar-water mixture matches with that of the silicone oil. A separate batch of sugar-water mixture was prepared for every batch of silicone oil. The properties of each liquid are given in Table \ref{tab:properties_fluids} and the properties of the liquid-liquid combination are given in Table \ref{tab:properties_combinations}. Since the density of the sugar-water mixture was always higher than that of the silicone oil, fluid-1 (upper fluid) was always silicone oil and fluid-2 was the refractive-index-matched sugar-water mixture. Since the glass slide was always made to slide downwards, silicone oil becomes the displaced (receding) fluid and the sugar-water mixture becomes the displacing (advancing) fluid. The viscosity ratio, defined as $\lambda = \mu_1/\mu_2$ was always found to be higher than one in the present study.

While matching the refractive index is the best way to conduct PIV experiments in both liquids, the interface disappears from direct visual sight. As a result, the interface position is determined by carefully tracing the particle paths as described below. To obtain the vector fields from particle images, preprocessing of the images is performed including average background subtraction and image equalization. Particle images are processed using a multi-grid, window-deforming PIV algorithm. The signal-to-noise ratio was improved using an ensemble PIV correlation in an interrogation window of 8 pixels $\times$ 8 pixels. The interface shapes were extracted by analyzing particle paths and the interface position was determined with the help of streakline image obtained by combining about 500 consecutive images. This technique works only when the flow is strictly two-dimensional and steady as is the case in the present study.

% \subsection{Cleaning procedure}
A cleaning protocol was strictly followed in all experiments to avoid any contamination. Before starting the experiments, the tank and glass slides were cleaned with isopropyl alcohol, followed by a thorough rinse with distilled water, and dried using a hot air gun to avoid any potential contamination. To control the contact angle, either hydrophilic or hydrophobic coatings were applied on the glass surface. After allowing the coating to dry, the slides were thoroughly rinsed in distilled water to remove any loose particles from the coating and then dried. Experiments were conducted only after the glass slides reached ambient temperatures to prevent any thermal Marangoni effects.

%%%%%%%%%%%%%%%%%%%%%%%%%%%%%%
%%%%%%%%%%%%%%%%%%%%%%%%%%%%%%
\begin{table}[h]
\centering % instead of \begin{center}
\caption{Operating parameters}
\begin{tabular}{ c | c  c  c c c} 
\hline
      Fluid combination &  
      \begin{tabular}{@{}c@{}}$U$(mm/s) \\ \end{tabular}  &
      \begin{tabular}{@{}c@{}}$Re$ \\ \end{tabular}  &
      \begin{tabular}{@{}c@{}}$Ca$ \\ \end{tabular}  &
       \begin{tabular}{@{}c@{}}Contact Angle \\ (deg.)\end{tabular}  & 
      \begin{tabular}{@{}c@{}} Surface \\ property \\ \end{tabular}  \\
\hline
\multirow{4}{4cm}{Combination-1: 5 cSt silicone oil (fluid-1) \& SW-1 (fluid-2)} & 0.5 &$4.77 \times 10^{-1}$ &  $6.48 \times 10^{-5}$ & $\approx 101.3$& Uncoated \\ 
 & 1 &$9.53 \times 10^{-1}$ & $1.30 \times 10^{-4}$ & $\approx 122.6$ & Coated\\ 
 & 1 &$9.53 \times 10^{-1}$ &  $1.30 \times 10^{-4}$ &$\approx 104.4$& Uncoated\\ 
 & 5 &$4.77 $ & $6.48 \times 10^{-4}$ & $\approx 110.2$& Uncoated\\
 \hline
\multirow{4}{4cm}{Combination-2: 10 cSt silicone oil (fluid-1) \& SW-2 (fluid-2)} & 0.5 & $4.77 \times 10^{-1}$ & $6.61 \times 10^{-5}$ & $\approx 62.8$ & Coated\\ 
 & 1 &$9.54 \times 10^{-1}$ &  $1.32 \times 10^{-4}$ & $\approx 106.6$ & Coated\\ 
 & 5 &$4.77$ & $1.19 \times 10^{-3}$ & $\approx 115.1$ & Coated\\ 
  & 5 &$4.77$ & $1.19 \times 10^{-3}$ & $\approx 117.5$ & Uncoated\\ 
 \hline
\multirow{3}{4cm}{Combination-3: 50 cSt silicone oil (fluid-1) \& SW-2 (fluid-2)}& 0.15 & $1.52 \times 10^{-1}$ & $2.01 \times 10^{-5}$ & $\approx132.1$ & Coated\\ 
  & 0.5 & $5.07 \times 10^{-1}$ & $6.71 \times 10^{-5}$ & $\approx140.8$ & Coated\\ 
 & 1 & $1.01$ & $1.34 \times 10^{-4}$ & $\approx148.5$ & Coated \\
 \hline
\label{tab:operating_parameters}
\end{tabular}
\begin{flushleft}
\begin{tablenotes}
      \footnotesize
      \item Uncoated represents a clean glass surface
      \item Coated represents a coated surface with hydrophilic coating
    \end{tablenotes}
\end{flushleft}
\end{table}

%%%%%%%%%%%%%%%%%%%%%%%%%%%
%%%%%%%%%%%%%%%%%%%%%%%%%%%%

%%%%%%%%%%%%%%%%%%%%%%%%%%%
\begin{figure}
\centering
\label{fig:two_phase_streakline_image}
\includegraphics[trim = 0mm 0mm 0mm 0mm, clip, angle=0,width=0.85\textwidth]{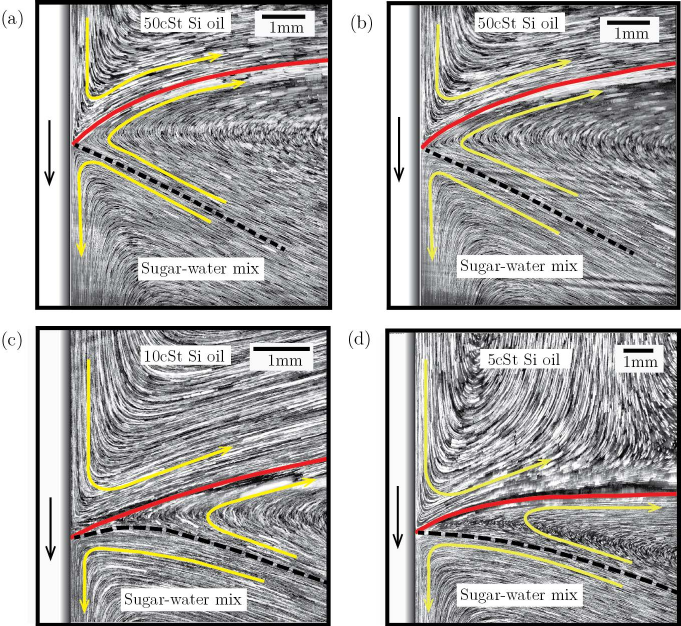}
\caption{Experimental streakline images showing the flow patterns for different liquid-liquid systems ($\lambda \approx 9.4$ for panels a and b; $\lambda \approx 1.8$ for panel c and $\lambda \approx 1$ for panel d).  The red solid line represents the interface between the silicone oil and sugar-water phases. The yellow arrow represents the direction of the flow in both phases. The black dashed line indicates the split-streamline in the sugar-water phase.}
\end{figure} 
%%%%%%%%%%%%%%%%%%%%%%
%%%%%%%%%%%%%%%%%%%%%%

%%%%%%%%%%%%%%%%%%%%%%%%%%%%
%%%%%%%%%%%%%%%%%%%%%%%%%%%
\begin{figure}
\centering
\subfigure[]
{\label{fig:geometry_huh}
\includegraphics[trim = 0mm 0mm 0mm 0mm, clip, angle=0,width=0.3\textwidth]{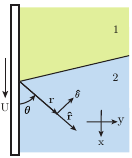}}
\hspace{3mm}
\subfigure[]
{\label{fig:rolling_flow_schematic}
\includegraphics[trim = 0mm 0mm 0mm 0mm, clip, angle=0,width=0.22\textwidth]{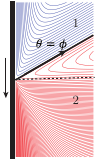}}
\hspace{3mm}
\subfigure[]{
\label{fig:coordinate_system}
  \includegraphics[trim = 0mm 0mm 0mm 0mm, clip, angle=0,width=0.3\textwidth]{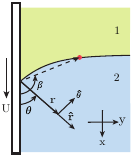}}
\caption{ (a) Cylindrical coordinate system ($r-\theta$) for the flow in a wedge with origin $(0,0)$ at the contact line based on HS71 theory \cite{huh1971hydrodynamic}. The moving plate advances from fluid-1 to fluid-2 with a speed $U$. (b) A typical flow pattern showing rolling motion in fluid-1 and split streamline in fluid-2 with material points along the interface ($\theta = \phi$) moving away from the contact line. (c) Coordinate system for local flow near a moving contact line with a curved wedge with $\beta(r)$ varying along the curved interface.}

\end{figure} 
%%%%%%%%%%%%%%%%%%%%%%
%%%%%%%%%%%%%%%%%%%%%%
The speed of the glass slide was independently varied to achieve various Reynolds and capillary numbers as given in Table \ref{tab:operating_parameters}. The Reynolds and capillary numbers are defined in terms of the lower fluid, i.e. sugar-water mixture, the interfacial tension, and the capillary length. This is consistent with liquid-gas experiments \cite{chen1997velocity,gupta2023experimental} where $Re$ and $Ca$ are also defined using the properties of the lower fluid. A phase plot showing all the experimental data points is shown in figure \ref{fig:two_phase_plot}. Figure \ref{fig:two_phase_streakline_image} shows streakline images with a silicone oil/sugar-water/glass system. 
The streakline image is obtained by combining the 500 particle images spanned over 2 seconds of real-time. Only those particle images are considered which are captured after reaching a steady state. 

\section{\label{sec:theory_background}Brief review of theoretical models}
The study of HS71 examined the dynamics near a moving contact line using a flat interface as shown schematically in figure \ref{fig:geometry_huh}. Since curved interfaces occur in experiments, the fixed wedge angle, $\alpha$, can simply be replaced by the variable local angle $\beta$  \cite{cox1986dynamics,chen1997velocity,gupta2023experimental}. Below, we briefly provide the analytical forms of the streamfunction as well as the interface shape based on state-of-the-art theoretical models.
%%%%%%%%%%%%%%%%%%%%%%%%%%%%%%%%%%%%%
%%%%%%%%%%%%%%%%%%%%%%%%%%%%%%%%%%%%%
%%%%%%%%%%%%%%%%%%%%%%%%%%%%%%%%%%%%%

\subsection{Models for flow fields} \label{sec:models_flow_fields}
In the viscous limit, the two-dimensional Stokes equations reduce to the biharmonic equation for the streamfunction, $\psi$,
\begin{equation}
    \nabla^4 \psi = 0.
\end{equation}
The analytical study of HS71 obtained a solution for the streamfunction in both fluids in terms of eight coefficients. The general solution of the streamfunction assumes the form
\begin{align}\label{eq:HS71}
\psi_{1,2} =& r(A_{1,2}(\phi)\sin\theta + B_{1,2}(\phi)\cos\theta + C_{1,2}(\phi)\theta \sin\theta + D_{1,2}(\phi)\theta \cos\theta), 
\end{align}
where subscripts 1 and 2 denote the fluid phases. The expressions for all the coefficients are given in Appendix \ref{secA1}. In the HS71 formulations, the interfacial velocity, $v_i^{HS}$, is equal to the radial velocity at the interface and is independent of the radial location along the interface given by
%%%%%%%%%%%%%%%%%%%%%%%%%%%
\begin{equation}
    v_i^{HS} = v_r(r,\phi) = U \left(\frac{(\phi \cos \phi - \sin \phi)(\sin^2\phi - \delta^2) + \lambda(\delta \cos \phi - \sin \phi)(\phi^2 - \sin^2\phi)}{\Delta}\right),
    \label{eq:Scriven2}
\end{equation}
%%%%%%%%%%%%%%%%%%%%%%%%%%%
To facilitate the comparison of the streamfunction with experiments where the interface is always curved, we modify the streamfunction in equn. \eqref{eq:HS71} for a curved interface by replacing the fixed angle $\phi$ with interface angle $\beta(r)$ (see figure \ref{fig:coordinate_system}). The local interface angle, $\beta(r)$, can be obtained from another model or directly measured in experiments. The streamfunction of a curved interface now becomes
%%%%%%%%%%%%%%%%%%%%%%%%%%%%%%%%%%%%%%%%%%%%%%%%%%%
%%%%%%%%%%%%%%%%%%%%%%%%%%%%%%%%%%%%%%%%%%%%%%%%%%%
%%%%%%%%%%%%%%%%%%%%%%%%%%%%%%%%%%%%%%%%%%%%%%%%%%%
%%%%%%%%%%%%%%%%%%%%%%%%%%%%%%%%%%%%%%%%%%%%%%%%%%%
%%%%%%%%%%%%%%%%%%%%%%%%%%%%%%%%%%%%%%%%%%%%%%%%%%%
%%%%%%%%%%%%%%%%%%%%%%%%%%%%%%%%%%%%%%%%%%%%%%%%%%%
%%%%%%%%%%%%%%%%%%%%%%%%%%%%%%%%%%%%%%%%%%%%%%%%%%%
%%%%%%%%%%%%%%%%%%%%%%%%%%%%%%%%%%%%%%%%%%%%%%%%%%%
%%%%%%%%%%%%%%%%%%%%%%%%%%%%%%%%%%%%%%%%%%%%%%%%%%%
\begin{equation}
\psi_{1,2} = r(A_{1,2}(\beta)\sin\theta + B_{1,2}(\beta)\cos\theta + C_{1,2}(\beta)\theta \sin\theta + D_{1,2}(\beta)\theta \cos\theta). 
\label{eq:psi_MWS_12}
\end{equation}
%%%%%%%%%%%%%%%%%%%%%%%%%%%%%%
%%%%%%%%%%%%%%%%%%%%%%%%%%%%%%
This solution is referred to as the modulated wedge solution (MWS) \cite{chen1997velocity} and was shown to make excellent predictions for the flow fields when $\lambda < 1$ \cite{gupta2023experimental}. 
The radial and tangential velocities can also be easily computed as
%%%%%%%%%%%%%%%%%%%%%%%%%%
\begin{eqnarray}\label{eq:MWS_vr}
   && v_r(r,\theta;\beta) = \frac{\partial f}{\partial\theta}, \\
   && \label{eq:MWS_vtheta} v_{\theta}(r,\theta;\beta) = -f - r\frac{\partial f}{\partial \beta}\frac{\partial \beta}{\partial r},
\end{eqnarray}
%%%%%%%%%%%%%%%%%%%%%%%%%%
%%%%%%%%%%%%%%%%%%%%%%%%%
Unlike the HS71 solution, the interfacial velocity now depends on both the radial and the angular velocities. Setting $\theta = \beta$ in equns. \eqref{eq:MWS_vr} and \eqref{eq:MWS_vtheta}, we get
%%%%%%%%%%%%%%%%%%%%%%%%%
\begin{equation}\label{eq:modulated_interfacial_speed}
    v_i^{MW} = v_r(r,\beta) \cos(\alpha-\beta) + v_{\theta}(r,\beta) \sin(\alpha-\beta).
\end{equation}
%%%%%%%%%%%%%%%%%%%%%%%%%
%%%%%%%%%%%%%%%%%%%%%%%%%
where $\alpha(r)$ is a measure of the slope of the interface. In the next section, we apply analytical models and show how interface shapes from theoretical models compare with experiments.

%%%%%%%%%%%%%%%%%%%%%%%%%%%%%%%%%%%%%%%%%%%%%%%%%%%
%%%%%%%%%%%%%%%%%%%%%%%%%%%%%%%%%%%%%%%%%%%%%%%%%%%
%%%%%%%%%%%%%%%%%%%%%%%%%%%%%%%%%%%%%%%%%%%%%%%%%%%
%%%%%%%%%%%%%%%%%%%%%%%%%%%%%%%%%%%%%%%%%%%%%%%%%%%
%%%%%%%%%%%%%%%%%%%%%%%%%%%%%%%%%%%%%%%%%%%%%%%%%%%
%%%%%%%%%%%%%%%%%%%%%%%%%%%%%%%%%%%%%%%%%%%%%%%%%%%
%%%%%%%%%%%%%%%%%%%%%%%%%%%%%%%%%%%%%%%%%%%%%%%%%%%
%%%%%%%%%%%%%%%%%%%%%%%%%%%%%%%%%%%%%%%%%%%%%%%%%%%
%%%%%%%%%%%%%%%%%%%%%%%%%%%%%%%%%%%%%%%%%%%%%%%%%%%
\subsection{Models for complete interface shape}\label{sec:Interface_Shape_theory}

In the classical work of Cox \cite{cox1986dynamics}, the region in the vicinity of a moving contact line was divided into three regions. The inner region, closest to the contact line was dominated by slip whereas the outer region was determined by the geometry of the problem under consideration. These two regions were matched together using the intermediate region. The well-known DRG model by Dussan \emph{et al.} \cite{dussan1991} proposed a composite solution for predicting the interface shape in the intermediate region by incorporating the static and Cox solutions to give:
 \begin{equation}\label{eq:drgcomp}
   \alpha(r) = g^{-1}\left[g(\omega_0)+Ca\ln\left(\frac{r}{l_c}\right)\right] + \left(\theta_s - \omega_0\right).
 \end{equation}
where 
\begin{equation}\label{eq:gtheta}
  g(x) = \int^x_0 \frac{\lambda((t^2 - \sin^2 t) ((\pi-t)+ \sin t \cos t)
    + ((\pi-t)^2- \sin^2 t) (t-\sin t \cos t))}{2 \sin t(\lambda^2  (t^2 -\sin^2 t
    )
    + 2 \lambda (t (\pi-t) + \sin^2 t)+ ((\pi-t)^2 -\sin^2 t))} dt.
\end{equation}
%%%%%%%%%%%%%%%%%%%%%%%%%%%%%%%%%%%
%%%%%%%%%%%%%%%%%%%%%%%%%%%%%%%%%%%
Here $\displaystyle \theta_{s}= f_0\left(\frac{r}{l_c};\omega_0\right)$ represents the static interface shape while $\omega_0$ is interpreted as an empirical parameter in the solution that can be determined by matching the solution with the interface shape from experiments. The expression of $f_0$ represents the two-dimensional static meniscus explained in more detail in a related study \cite{gupta2023experimental} and given in appendix \ref{secB}.  

%%%%%%%%%%%%%%%%%%%%%%%%%%%%%%%%
%%%%%%%%%%%%%%%%%%%%%%%%%%%%%%%%
%%%%%%%%%%%%%%%%%%%%%%%%%%%%%%%%

\section{Results}
\label{sec:results}
In the present study, the flow dynamics near a moving contact line are investigated mainly at $\lambda \geq 1$. The following subsections discuss the results on experimental interface shapes, flow fields, and interfacial speeds along with their comparisons with available theoretical models.   

%%%%%%%%%%%%%%%%%%%%%%%%%%%%%%%%%%
%%%%%%%%%%%%%%%%%%%%%%%%%%%%%%%%%%

\subsection{Interface shape}\label{sec:interface_shape}
\begin{figure}
\centering
\subfigure[]{\label{fig:10cst_shape}
\includegraphics[trim = 0mm 0mm 0mm 0mm, clip, angle=0,width=0.45\textwidth]{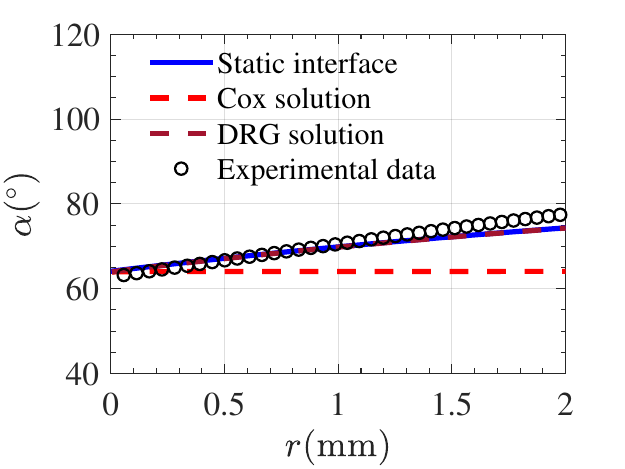}}
\subfigure[]{\label{fig:combined_shape}
\includegraphics[trim = 0mm 0mm 0mm 0mm, clip, angle=0,width=0.45\textwidth]{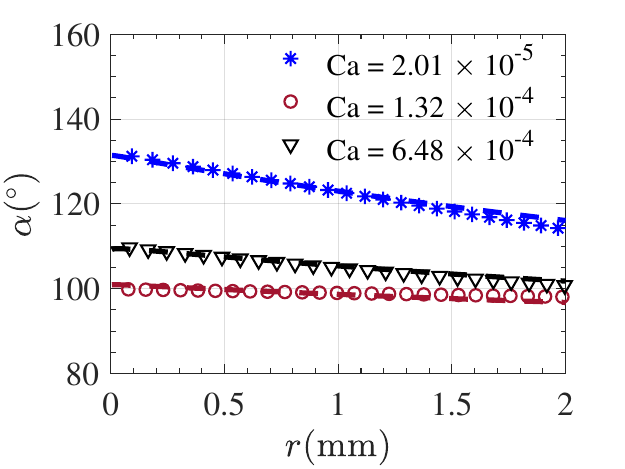}}
\caption{A comparison of experimental interface shape with static shape (blue solid curve), Cox model (red dashed curved), and DRG model (maroon dashed curve) for combination-2 with $Ca = 6.61\times 10^{-5}$. The solid surface is coated with a hydrophilic film and the DRG model results in $\omega_0 \approx 64.1^{\circ}$. (b) A comparison of interface shapes for three different combinations of liquids against the DRG model (dashed line): ~$\color{blue}\mathlarger{\mathlarger{\mathlarger{\ast}}}$: Combination-3 with $Ca=2.01\times10^{-5}$, $\omega_0 \approx 131.8^{\circ}$;~ $\color{black}\triangledown$: Combination-1 with $Ca=6.48\times10^{-4}$, $\omega_0 \approx 110.4^{\circ}$;~$\color{maroon}{\boldsymbol\circ}$: Combination-2 with $Ca=1.32\times10^{-4}$, $\omega_0 \approx 107.9^{\circ}$. The fluid combinations are described in Table \ref{tab:properties_combinations}.}
\label{fig:interface_shape_DRG_two_phase}
\end{figure}

Figure \ref{fig:interface_shape_DRG_two_phase} shows a comparison of interface shape predicted by DRG solution \cite{dussan1991} and the experiments. The matching procedure is outlined in figure \ref{fig:10cst_shape} where three distinct analytical curves are shown. The static interface shape corresponds to the `outer' solution characterised completely by the static contact angle. The Cox solution is the local solution valid very close to the contact line, and hence deviates from the experimental results away at distances beyond 0.3 mm from the contact line. The DRG solution, on the other hand, is a composite solution incorporating the Cox solution for small $r$ and the static solution for large $r$, where $r$ is the distance along the interface measured from the contact line. The DRG solution is parameterised in terms of a single parameter $\omega_0$ which is determined by fitting the equn. \ref{eq:drgcomp} to the experimental data. This yields $\omega_0 \approx 64.1^{\circ}$. 

Figure \ref{fig:combined_shape} shows a comparison of three distinct liquid-liquid systems given in table \ref{tab:properties_combinations} against the DRG model. The DRG model is found to predict the interface shape over the whole region of the flow with just a single fitting parameter. The interface shape extracted from these models and experiments can be used to determine the flow field in both the fluids using the MWS theory discussed in \S \ref{sec:theory_background} and given in equn. \ref{eq:psi_MWS_12}.

%%%%%%%%%%%%%%%%%%%%%%%%%%%%%%%%%%%%%%
%%%%%%%%%%%%%%%%%%%%%%%%%%%%%%%%%%%%%%
%%%%%%%%%%%%%%%%%%%%%%%%%%%%%%%%%%%%%%
%%%%%%%%%%%%%%%%%%%%%%%%%%%%%%%%%%%%%%

\subsection{Flow field}\label{sec:flow_fields_ex_MWS}
\begin{figure}
\centering
\subfigure[]{
\label{fig:streamline_plot_10cst_1mm}
\includegraphics[trim = 11mm 0mm 15mm 0mm, clip, angle=0,width=0.43\textwidth]{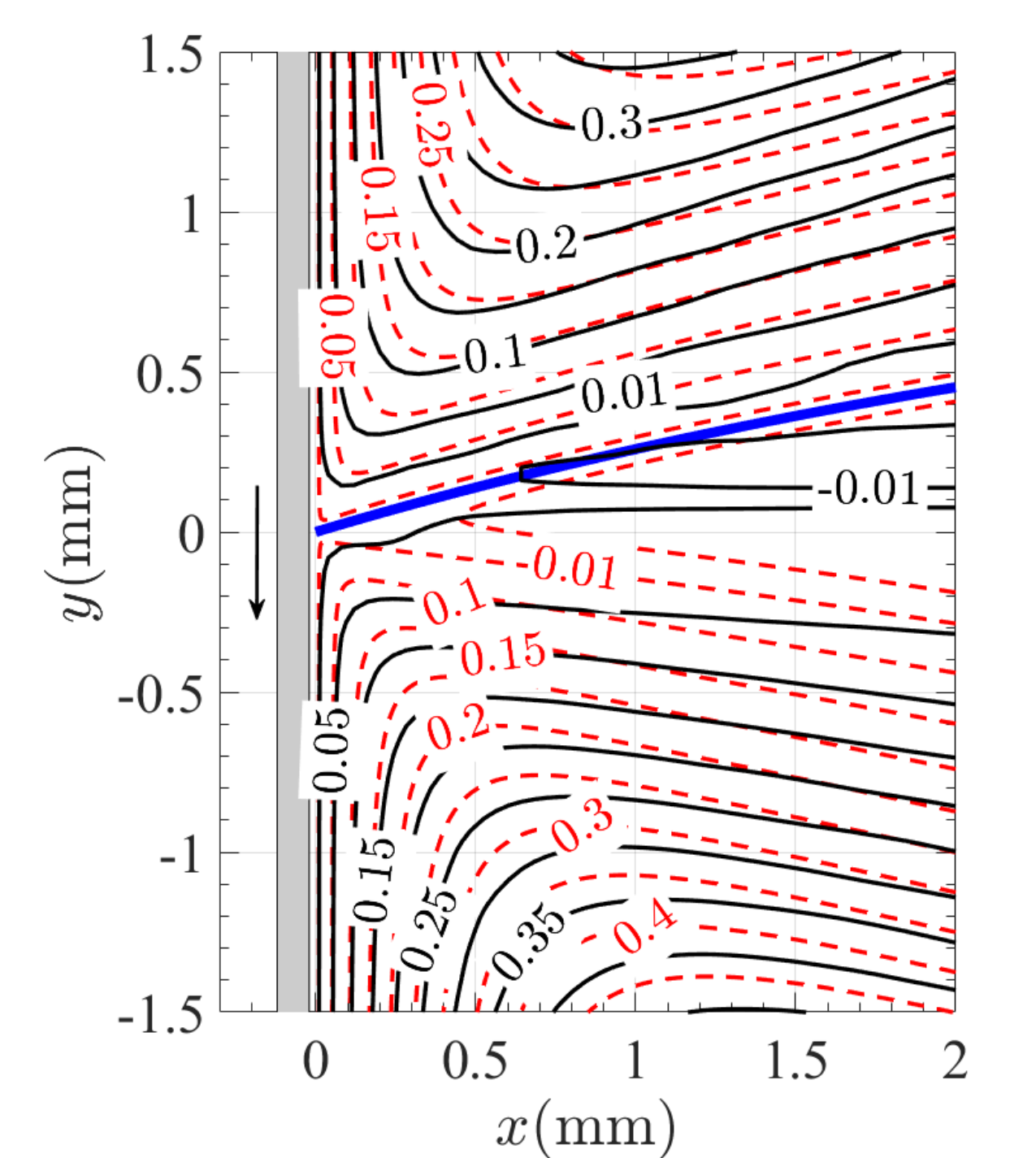}}
\hspace{10mm}
\subfigure[]{\label{fig:streamline_plot_10cst_5mm}
\includegraphics[trim = 12mm 0mm 13mm 0mm, clip, angle=0,width=0.43\textwidth]{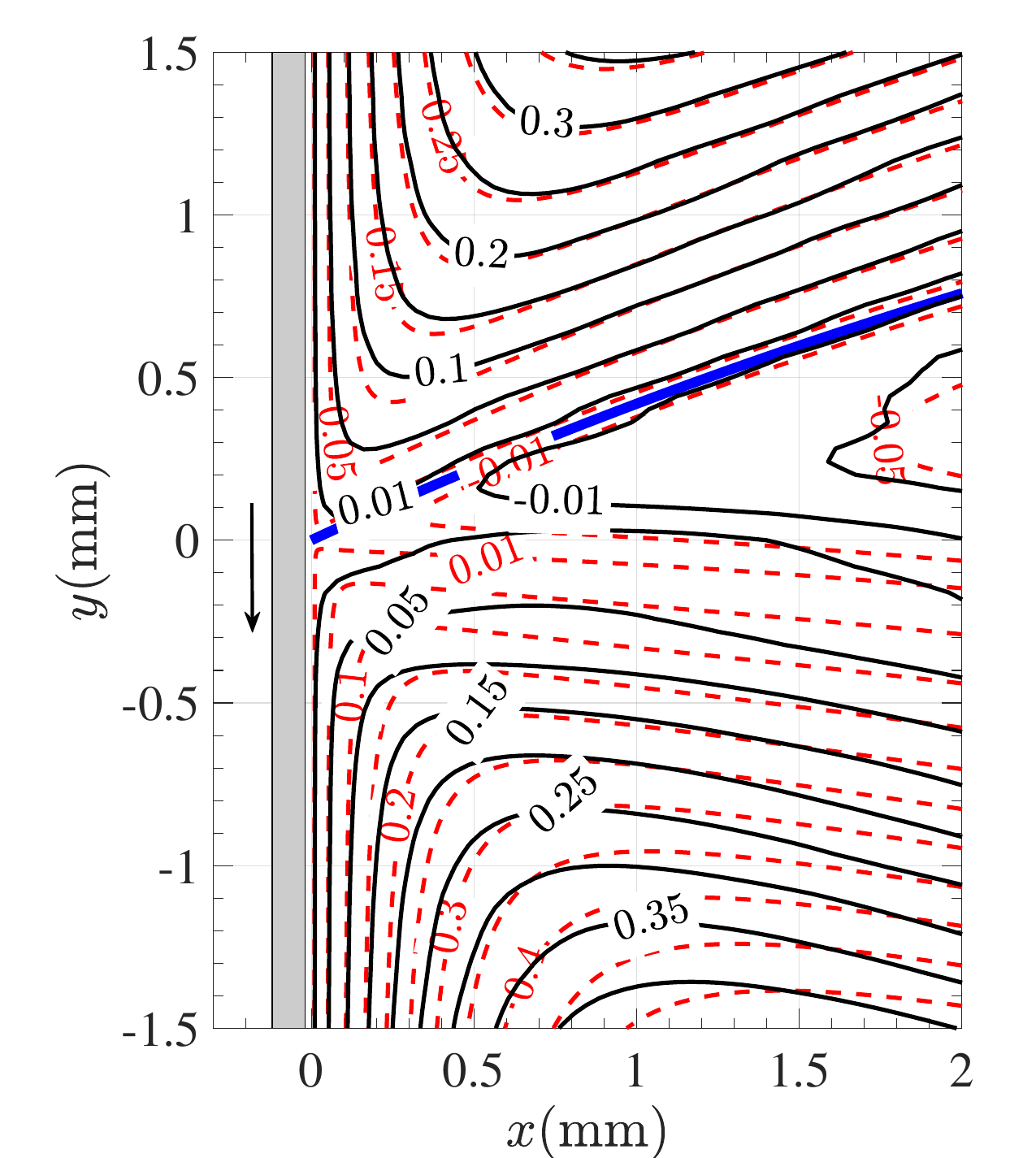}}
\caption{Contours of streamfunction obtained from experiments (black solid curves), MWS theory (red dashed curved) for (a) combination-2 at 
$Re= 9.54 \times 10^{-1}$, and $Ca = 1.32\times 10^{-4}$ (b) combination-2 at $Re= 4.77$, and $Ca = 6.61\times 10^{-4}$. The glass slide is coated with a hydrophilic coating. The blue solid curve represents the interface. The fluid combinations are described in Table \ref{tab:properties_combinations}.}
\label{fig:streamline_plot_10cst}
\end{figure}

The key goal of this study is to determine the flow fields near a moving contact line and compare the results against theoretical predictions. The velocity field, in the form of a vector field, is obtained from PIV data. The vector field is then used to determine streamfunction using the equations
\begin{equation}
    \frac{\partial \psi}{\partial y} = u, \quad \frac{\partial \psi}{\partial x} = -v,
\end{equation}
by setting $\psi=0$ on the moving wall. Figure \ref{fig:streamline_plot_10cst} shows a comparison between the experimentally determined streamfunction contours and theoretical predictions from MWS theory (equn. \ref{eq:psi_MWS_12}). Recall that the MWS theory is a direct extension of the HS71 theory with only a small modification involving the interface angle. The solid curves represent the experimental flow field whereas the dashed curve are obtained from theory. Given the complexity of the experiments, the agreement between theory and experiments is impressive. The HS71 theory is not only able to predict the overall shape of the flow field, but is able to make a quantitative prediction for the streamfunction field in the vicinity of the contact line. The experiments also confirm the presence of split-streamline and its location in the less viscous fluid just as predicted by HS71 theory and shown in figure \ref{fig:rolling_flow_schematic}. It is important to note that no adjustable parameters were employed in the theory. To the best of our knowledge, such a detailed and direct comparison between HS71 theory and experiments in a liquid-liquid moving contact line has never been carried out before.

%%%%%%%%%%%%%%%%%%%%%%%%
%%%%%%%%%%%%%%%%%%%%%%%%
%%%%%%%%%%%%%%%%%%%%%%%%
%%%%%%%%%%%%%%%%%%%%%%%%
%%%%%%%%%%%%%%%%%%%%%%%%
%%%%%%%%%%%%%%%%%%%%%%%%

\subsection{Interfacial speed}\label{sec:interfacial_speed}
Interfacial speed ($v_i$) is a tangential speed measured along the interface by projecting the experimentally determined flow field onto the interface. The direction of the interface velocity, determined from the sign of $v_i$, reveals the flow pattern in both the fluids. In the present study, since the glass slide is advancing downwards from silicone oil to sugar-water mixture, a positive value of $v_i$ corresponds to rolling motion in the upper (silicone oil) fluid and split-streamline motion in the lower (sugar-water mixture) fluid. 

\begin{figure}
\centering
\subfigure[]{
\label{fig:interface_speed_two_fluid}
\includegraphics[trim = 2mm 0mm 8mm 4mm, clip, angle=0,width=0.47\textwidth]{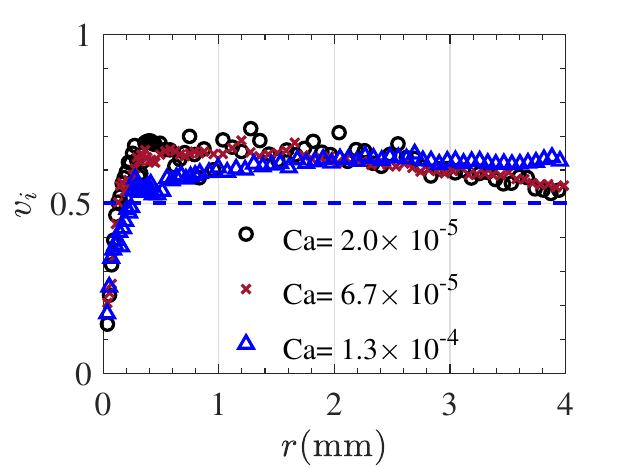}
}
\subfigure[]{
\label{fig:slip_velocity_two_fluid}
\includegraphics[trim = 2mm 0mm 8mm 4mm, clip, angle=0,width=0.47\textwidth]{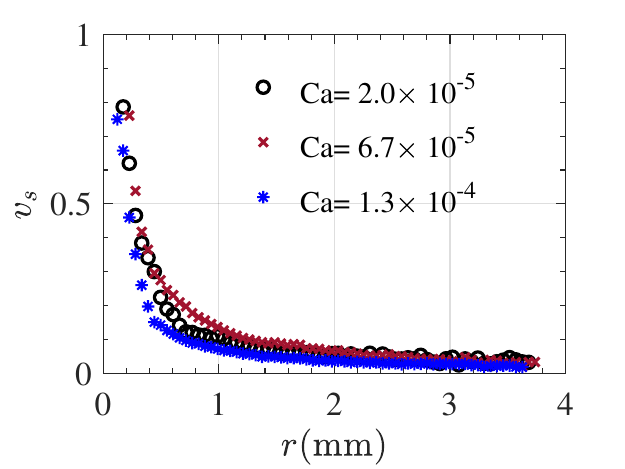}
}
\caption{(a) A comparison of experimental interfacial speed, $v_i$ (measured along the interface non-dimensionalized with the speed of the plate) against HS71 theory. (b) Slip velocities (non-dimensionalized with the speed of the plate) measured in fluid-2. The data corresponds to combination-3 given in table \ref{tab:properties_combinations}
performed over the solid surface coated with hydrophilic coating.}
\label{fig:interfacial_speed_comparison_10cst_Si_two_phase}
\end{figure}

Figure \ref{fig:interfacial_speed_comparison_10cst_Si_two_phase} shows the interfacial speed data for 50 cSt silicone oil-sugar water mixture system. The positive value of speed indicates that fluid particles at the interface are moving away from the contact line. This is clearly consistent with the rolling motion in fluid-1 and split-streamline motion in fluid-2. The HS71 theory correctly predicts the sign of the tangential velocity, but there is clearly deviation from experimental observations. 

The singularity in the HS71 theory also manifests in the tangential speed given in equn. \ref{eq:Scriven2}, where the speed is independent of radial location from the contact line. As per HS71 theory, material points move at a constant speed along the interface since theoretical speed, $v_i^{HS}$, is independent of $r$. This indicates that material points approaching the interface either from fluid-1 or fluid-2 instantaneously speed up (i.e. possess infinite acceleration) and then move along the interface. The experiments clearly offer a resolution to the singularity as shown in figure \ref{fig:interface_speed_two_fluid}. The interfacial speed is clearly shown to rapidly decrease as the contact line is approached. Hence material points gradually accelerate and attain a constant speed as they move away from the contact line.

Another important finding of the present study is the presence of slip near the moving wall in the vicinity of the contact line as shown in figure \ref{fig:slip_velocity_two_fluid}. Here, $r=0$ represents the contact line and $r$ increases as one moves away from the contact line in fluid-2. The slip velocity is defined as the difference between the fluid and plate velocity at the moving wall normalised by the plate velocity. A value of $v_s=1$ represents perfect slip and $v_s=0$ represents no-slip. At a distance of about 1 mm away from the contact line, the no-slip boundary condition appears to be well enforced. But as one approaches the contact line, the slip velocity rapidly increases. This is another key finding of the study and offers a pathway for the resolution of the contact line singularity. 

%%%%%%%%%%%%%%%%%%%%%%%%
%%%%%%%%%%%%%%%%%%%%%%%%

%%%%%%%%%%%%%%%%%%%%%%%%
%%%%%%%%%%%%%%%%%%%%%%%%

\section{Conclusion}
\label{sec:conclusion}
In this study, the velocity in the vicinity of a moving contact line in a liquid/liquid/solid system is investigated using particle image velocimetry experiments. By matching the refractive-index in both liquids, the flow field in both liquids is simultaneously obtained in a single experiment. A quantitative comparison is made with theoretical models and it is shown that the theory of HS71 makes excellent predictions for flow in both liquids. To facilitate comparison between experiments and HS71 theory, the shape of the interface is determined and provided as an input to the theory which results in the modulated wedge solution (MWS). HS71 theory is also shown to make reasonable predictions for the interfacial speed. For the interface shape, the DRG model was found to make a good prediction. More sophisticated models for interface shape do exist, but these models are largely valid in the limit of $\lambda \rightarrow 0$, i.e., in the free-surface limit.

A key finding of the study is how the singularity in HS71 theory is arrested in the experiments. Unlike the HS71 theory, experiments clearly reveal that interface velocity rapidly varies in the vicinity of the moving contact line. The absence of singularity is also consistent with the presence of slip velocity along the moving wall in the vicinity of the moving contact line.

Since the viscosity ratio in the present study is close to unity, we believe the findings presented in this study will be easily amenable to numerical investigation. We hope the present study will spur new theoretical and numerical studies in liquid/liquid/solid moving contact lines.

\backmatter

%\bmhead{Supplementary information}

%If your article has accompanying supplementary file/s please state so here. 

\bmhead{Acknowledgments}

This work was supported by the Science and Engineering Research Board, Department of Science and Technology (DST), India  (Grant number: CRG/2021/007096). The authors would also like to express their gratitude for the support received from the Fund for Improvement of Science and Technology Infrastructure (FIST) of India (Grant Number: SR/FST/ETI-397/2015/(C)), and the JICA Friendship Project.

% Please refer to Journal-level guidance for any specific requirements.

\section*{Declarations}

\begin{itemize}
%\item Funding:
\item Conflict of interest/Competing interests: 
The authors have no competing interests to declare that are relevant to the content of this article.
%\item Ethics approval 
%\item Consent for publication
\item Availability of data and materials:
The datasets generated during and analysed during the current study are available from the corresponding author upon reasonable request.
%\item Authors' contributions
\end{itemize}

\noindent
%If any of the sections are not relevant to your manuscript, please include the heading and write `Not applicable' for that section. 

% \bigskip

% Springer journals and proceedings: \url{https://www.springer.com/gp/editorial-policies}

% \end{flushleft}

\begin{appendices}

\section{A brief description of the formulation of streamfunction in a fixed wedge}\label{secA1}
The governing equation near a moving contact line for a fixed wedge is given as:
\begin{equation}
    \nabla^4 \psi = 0,
\end{equation}
where $\psi$ can be expanded as follows:
\begin{align}
\psi_{1,2} =& r(A_{1,2}(\phi)\sin\theta + B_{1,2}(\phi)\cos\theta + C_{1,2}(\phi)\theta \sin\theta + D_{1,2}(\phi)\theta \cos\theta), 
\end{align}
%%%%%%%%%%%%%%%%%%%%%%%%%%%%
%%%%%%%%%%%%%%%%%%%%%%%%%%%
The eight coefficients in the above equation are determined by applying standard boundary conditions on the moving wall (no-slip) and along the interface (continuity of tangential velocity and continuity of tangential stress). 
\begin{align*}
A_1 =& \frac{1}{\sin\phi}(-C_1\phi\sin\phi - D_1\delta\cos\phi) && \\[5pt]
B_1 =&  - D_1\pi \\[5pt]
C_1 =& \frac{1}{\delta}\left(U - D_1(\delta\frac{\cos\phi}{\sin\phi}-1)\right) \\[5pt]
D_1 =& \frac{U\sin\phi\cos\phi}{\Delta}\left((\sin^2\phi-\delta\phi)+ \lambda(\phi^2 - \sin^2\phi)- \pi \tan\phi \right) \\[5pt]
A_2 =& \frac{1}{\sin\phi}(-C_2\phi\sin\phi - D_2\phi\cos\phi) \\[5pt]
B_2 =& 0  \\[5pt]
C_2 =& \frac{1}{\phi}\left(U + D_2(1-\phi\frac{\cos\phi}{\sin\phi})\right) \\[5pt]
D_2 =& \frac{U\sin\phi\cos\phi}{\Delta}\left(\sin^2\phi - \delta^2 + \lambda(\delta\phi - \sin^2\phi)- \lambda\pi \tan\phi \right) 
\end{align*}
where $\Delta= (\sin\phi\cos\phi-\phi)(\delta^2- \sin^2\phi) + \lambda(\phi^2-\sin^2\phi)(\delta - \sin\phi\cos\phi)$ and $\delta = \phi - \pi$.

%%%%%%%%%%%%%%%%%%%%%%%%%%%%%%%%%%%%%%%%%%%%%
%%%%%%%%%%%%%%%%%%%%%%%%%%%%%%%%%%%%%%%%%%%%%
%%%%%%%%%%%%%%%%%%%%%%%%%%%%%%%%%%%%%%%%%%%%%
%%%%%%%%%%%%%%%%%%%%%%%%%%%%%%%%%%%%%%%%%%%%%
%%%%%%%%%%%%%%%%%%%%%%%%%%%%%%%%%%%%%%%%%%%%%
\section{Static interface shape in the DRG model}\label{secB}
The static interface shape $\theta_s$ can be described using  horizontal and vertical coordinates $H_s$ and $X$ with the origin at the contact line as
%%%%%%%%%%%%%%%%%%
\begin{equation}
    % f_0 = \frac{\pi}{2} + \tan^{-1}\left(\frac{dh_s}{dx_s}\right)
    \theta_s = \frac{\pi}{2} - \tan^{-1}\left(\frac{dH_s}{dX}\right),
  \label{eq:f0}
\end{equation}
%%%%%%%%%%%%%%%%%%%%%%
In the above expression, the static solution is represented by subscript $s$ and $X \in [0,X_0]$.
A schematic for static interface shape can be seen in figure~\ref{fig:f0men} (a dashed curve)
 Using the solution of the non-linear Young-Laplace equation, $H_s(X)$ can be written as follows:
%%%%%%%%%%%%%%%%%%%%%%%%%%%%%%%%%%
\begin{figure}
\centering
\includegraphics[trim = 0mm 0mm 0mm 0mm, clip, angle=0,width=0.3\textwidth]{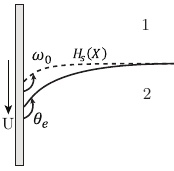}
\caption{A schematic shows a plate moving with speed $U$ with respect to an interface of two immiscible fluids. The interface touches the wall at the contact line with an angle assumed to be the same as the equilibrium contact angle $\theta_e$. $H_s(X)$ represents the static shape obtained by imposing a contact angle of $\omega_0$ at the wall.}
\label{fig:f0men}
\end{figure}
%%%%%%%%%%%%%%%%%%%%%%%%%%%%%%%%%%%
\begin{equation}
\nonumber
  H_s(X) = \cosh^{-1}\left(\frac{2}{X_0-X}\right) - \cosh^{-1}\left(\frac{2}{X_0}\right) - \left(4 - (X_0-X)^2\right)^{1/2} + \left(4 - X_0^2\right)^{1/2},
  \label{eq:static_shape_parametric}
\end{equation}
where
\begin{equation} 
\nonumber
  X_0 = \sqrt{2}\left(1-\sin\omega_0\right)^{1/2}. 
\end{equation}
%%%%%%%%%%%
In this context, $\omega_0$ denotes the angle at the contact line ($x=0$). Instead of being explicitly prescribed, it is determined through an iterative process that minimizes the root mean square error between the experimental interface shape and the interface shape predicted by the DRG model in equn.\eqref{eq:drgcomp}. 
%%%%%%%%%%%%%%%%%%%%%%%%%%%%%%
\end{appendices}

\bibliographystyle{unsrt}
\bibliography{bibfile}% common bib file

\end{document}